\documentclass[aip,reprint,a4paper,floatfix,amsmath,amssymb,amsfonts,noshowpacs]{revtex4-1}
\usepackage{newtxtext,newtxmath}
\usepackage[utf8]{inputenc}
\usepackage{graphicx}
\usepackage[dvipsnames]{xcolor}
\usepackage{chemist}
\usepackage{algorithm2e}
\usepackage[pdftex,unicode=true,bookmarks=false,breaklinks=false,pdfborder={0 0 1},colorlinks=true]{hyperref}
\hypersetup{linkcolor=blue,citecolor=blue,urlcolor=blue}
\usepackage[textwidth=17.5cm,textheight=23.5cm,verbose,pdftex]{geometry}
\usepackage{soul}
\usepackage{tikz}
\usepackage{chemfig}
\usepackage{mhchem}
\usetikzlibrary{calc,fadings,decorations.markings}
 
\DeclareUnicodeCharacter{03B1}{$\alpha$}
\DeclareUnicodeCharacter{03B2}{$\beta$}
\DeclareUnicodeCharacter{03B5}{$\epsilon$}
\DeclareUnicodeCharacter{03BA}{$\kappa$}

\newcommand{\gox}{Ga$_2$O$_3$}
\newcommand{\agox}{$\upalpha$-Ga$_2$O$_3$}
\newcommand{\bgox}{$\upbeta$-Ga$_2$O$_3$}
\newcommand{\aaox}{$\upalpha$-Al$_2$O$_3$}
\newcommand{\aox}{Al$_2$O$_3$}
\newcommand{\algox}{(Al,Ga)$_2$O$_3$}

\newcommand{\algoxcon}{(Al$_x$Ga$_{1-x}$)$_2$O$_3$}
\newcommand{\aingox}{$\upalpha$-(In,Ga)$_2$O$_3$}
\newcommand{\ingoxcon}{(In$_x$Ga$_{1-x}$)$_2$O$_3$}

\newcommand{\bgoxmitin}{$\upbeta$-(In)Ga$_2$O$_3$}
\newcommand{\ino}{In$_2$O$_3$}

\newcommand{\oxflux}{$\phi_{\text{O}}$}
\newcommand{\rme}{$R_{\text{Me}}$} 
\newcommand{\ogaflux}{$R_{\text{O}}$} 
\newcommand{\ogafluxdef}{$\phi_{\text{O}}/\phi_{\text{Ga}}$}
\newcommand{\influx}{$\phi$\textsubscript{In}}
\newcommand{\gaflux}{$\phi$\textsubscript{Ga}}
\newcommand{\ingaflux}{$\phi_{\text{In}}/\phi_{\text{Ga}}$}
\newcommand{\ingafluxabb}{$R_{\text{In}}$}

\newcommand{\xrdqw}{2$\uptheta$-$\upomega$}
\newcommand{\fluxunit}{nm$^{-2}$s$^{-1}$}
\newcommand{\tg}{$T_{\text{G}}$}
\newcommand{\gr}{$\Gamma$}
\newcommand{\fga}{$\phi_{\text{Ga}}$}
\newcommand{\fin}{$\phi_{\text{In}}$}
\newcommand{\fo}{$\phi_{\text{O}}$}

\newcommand{\aalo}{$\upalpha$-Al$_2$O$_3$}
\newcommand{\agao}{$\upalpha$-Ga$_2$O$_3$}
\newcommand{\bgao}{$\upbeta$-Ga$_2$O$_3$}
\newcommand{\egao}{$\upepsilon$/$\upkappa$-Ga$_2$O$_3$}
\newcommand{\aingao}{$\upalpha$-(In$_x$Ga$_{1-x}$)$_2$O$_3$}
\newcommand{\aalgao}{$\upalpha$-(Al$_x$Ga$_{1-x}$)$_2$O$_3$}
\newcommand{\balgao}{$\upbeta$-(Al$_x$Ga$_{1-x}$)$_2$O$_3$}
\newcommand{\gso}{Ga$_2$O}

\newcommand{\aal}{$\upalpha$-Al$_2$O$_3$($10\bar{1}0$)}

\newcommand{\aga}{$\upalpha$-Ga$_2$O$_3$($10\bar{1}0$)}
\newcommand{\ainga}{$\upalpha$-(In$_x$Ga$_{1-x}$)$_2$O$_3$($10\bar{1}0$)}

\begin{document}

\preprint{AIP/123-QED}

\title{Growth, catalysis and faceting of \agao\ and \aingao\ on $m$-plane \aalo\ by molecular beam epitaxy}

\author{Martin S. Williams}
\email[Electronic mail: ]{marwilli@uni-bremen.de}
\altaffiliation{These authors contributed equally to this work.}
\affiliation{ 
Institute of Solid State Physics, University of Bremen, Otto-Hahn-Allee 1, 28359, Bremen, Germany
}%

\author{Manuel Alonso-Orts}
\email[Electronic mail: ]{alonsoor@uni-bremen.de}
\altaffiliation{These authors contributed equally to this work.}
\affiliation{ 
Institute of Solid State Physics, University of Bremen, Otto-Hahn-Allee 1, 28359, Bremen, Germany
}%
\affiliation{MAPEX Center for Materials and Processes, University of Bremen, Bibliothekstraße 1, 28359, Bremen,
Germany}

\author{Marco Schowalter}
\affiliation{ 
Institute of Solid State Physics, University of Bremen, Otto-Hahn-Allee 1, 28359, Bremen, Germany
}%

\author{Alexander Karg}
\affiliation{ 
Institute of Solid State Physics, University of Bremen, Otto-Hahn-Allee 1, 28359, Bremen, Germany
}%

\author{Sushma Raghuvansy}
\affiliation{ 
Institute of Solid State Physics, University of Bremen, Otto-Hahn-Allee 1, 28359, Bremen, Germany
}%

\author{Jon P. McCandless}
\affiliation{ 
School of Electrical and Computer Engineering, Cornell University, 229 Phillip's Hall, 14853, New York, United States of America
}%

\author{Debdeep Jena}
\affiliation{ 
School of Electrical and Computer Engineering, Cornell University, 229 Phillip's Hall, 14853, New York, United States of America
}%
\affiliation{ 
Department of Material Science and Engineering, Cornell University, Bard Hall, 14853, New York, United States of America
}%
\affiliation{
Kavli Institute at Cornell for Nanoscale Science, Cornell University, Ithaca, New York, NY-14853, United States of America
}%

\author{Andreas Rosenauer}

\author{Martin Eickhoff}
\affiliation{ 
Institute of Solid State Physics, University of Bremen, Otto-Hahn-Allee 1, 28359, Bremen, Germany
}%
\affiliation{MAPEX Center for Materials and Processes, University of Bremen, Bibliothekstraße 1, 28359, Bremen,
Germany}

\author{Patrick Vogt}
\email[Electronic mail: ]{pvogt@uni-bremen.de}
\affiliation{ 
Institute of Solid State Physics, University of Bremen, Otto-Hahn-Allee 1, 28359, Bremen, Germany
}%

\date{\today}

\begin{abstract}
The growth of \agao\ and \aingao\ on $m$-plane \aal\ by molecular beam epitaxy (MBE) and metal-oxide-catalyzed epitaxy (MOCATAXY) is investigated. By systematically exploring the parameter space accessed by MBE and MOCATAXY, phase-pure \aga\ and \ainga\ thin films are realized. The presence of In on the \agox\ growth surface remarkably expands its growth window far into the metal-rich flux regime and to higher growth temperatures. With increasing O-to-Ga flux ratio (\ogaflux), In incorporates into \aingao\ up to $x \leq 0.08$. Upon a critical thickness, $\upbeta$-\ingoxcon\ nucleates and subsequently heteroepitaxially grows on top of $\upalpha$-\ingoxcon\ facets. Metal-rich MOCATAXY growth conditions, where \agox\ would not conventionally stabilize, lead to single-crystalline \agox\ with negligible In incorporation and improved surface morphology.  Higher \tg\ further results in single-crystalline \agox\ with well-defined terraces and step edges at their surfaces. For \ogaflux\ $\leq$ 0.53, In acts as a surfactant on the \agox\ growth surface by favoring step edges, while for \ogaflux\ $\geq$ 0.8, In incorporates and leads to a-plane $\upalpha$-\ingoxcon\ faceting and the subsequent ($\bar{2}$01) $\upbeta$-\ingoxcon\ growth on top. Thin film analysis by STEM reveals highly crystalline \agox\ layers and interfaces. We provide a phase diagram to guide the MBE and MOCATAXY growth of single-crystalline \agox\ on \aal.  
\end{abstract}

\maketitle

\section{\label{sec:level1}Introduction}

The ultra-wide band gap semiconductor gallium oxide (\gox) has experienced tremendous interest for high-power electronics, whose development is essential to reduce energy loss in power converters \cite{Venkata2022}. Monoclinic ($\upbeta$-) \gox\ can be easily $n$-type doped by Sn, Si or Ge. The availability of commercial \bgox\ substrates reduces the material costs for device production compared to other wide band gap materials, such as SiC or GaN \cite{Reese2019}.

Other polymorphs of \gox\, such as the orthorhombic structure ($\upepsilon/\upkappa$-\gox) or the corundum structure ($\upalpha$-\gox), can also be epitaxially grown, with the latter being the polymorph with the widest band gap, of $E_{\text{g}} \approx 5.3\,\text{eV}$\cite{Jinno2021}, and isostructural to \aalo. This allows band gap engineering of \aalgao\ on \aalo\ over the whole composition range of $0 \leq x \leq 1$ \cite{Jinno2021} as the phase-stability of \aalgao\ on \aalo\ increases with increasing Al content \cite{McCandless2021}. The growth of high-quality \balgao\ thin films and the fabrication of high-electron-mobility transistors based on \balgao\ is limited to $x \lessapprox 0.3$ \cite{Ahmadi2017,Bhuiyan2020}. Those features of \agao\ provide an alternative route to develop high Al-mole fraction \algoxcon\ alloys for high power electronics.

The epitaxial growth of \agao\ and \aalgao\ on \aalo\ has been investigated by molecular beam epitaxy (MBE) \cite{Kracht2018,McCandless2021,Jinno2021,McCandless2023}, chemical vapor deposition (CVD)\cite{Dang2018,Uno2020} and pulsed laser deposition (PLD)\cite{Kneiss2021,Petersen2023}. In conventional MBE (hereafter named as `MBE'), i.e., by using elemental Ga and active O as source materials, the growth of \gox\ is limited by the formation and subsequent desorption of its volatile suboxide \gso\ and complex 2-step kinetics \cite{Vogt2018model}. This growth kinetics hampers the adsorption-controlled growth of \gox\ in the Ga-rich regime where its growth ceases in the excess of Ga adsorbates\cite{Vogt2016, Vogt2021}, thus, \gox\ is typically grown in the O-rich regime.

These growth features have a detrimental impact on the crystalline and transport properties of \gox\ thin films, for example, Ga vacancies ($V_{\text{Ga}}$) formed in \bgox\ when grown in the O-rich regime may act as compensating acceptors\cite{Lany2018}. To suppress the formation of $V_{\text{Ga}}$ and thus to potentially improve the electrical properties of \agao, its growth in the Ga-rich regime with improved crystalline quality is desirable\cite{Korhonen2015,McCandless2023}.

To overcome the detrimental growth kinetics of \gox\ occurring during MBE growth, a new MBE variant has been recently developed:~metal-oxide-catalyzed epitaxy (MOCATAXY) \cite{Vogt2018}. This growth method is based on metal-exchange catalysis (MEXCAT)\cite{Vogt2017,Kracht2017} and the use of the catalysts In, Sn, In$_2$O or SnO \cite{Vogt2022}. MOCATAXY expands the growth window of \bgao, \balgao, \agao, \egao\ and In$_2$O$_3$ deep into the metal-rich regimes and enables higher \tg\ while generally improving the properties, such as crystalline quality and surface roughness, of the thin films \cite{Vogt2017,Mazzolini2018,Vogt2018,Mauze2020,Mazzolini2020,Vogt2022,McCandless2023}. MOCATAXY, with In as a catalyst, emerges due to the favorable formation of intermediate and higher reaction efficiencies of \ino\ over \gox\ and the subsequent thermodynamically driven exchange of In-O bonds by Ga-O bonds \cite{Vogt2017}, and is mathematically explained for elemental and molecular catalysts in Ref.~\citenum{Vogt2022}.

To stabilize \agao\ on \aalo\ by MBE, the crystalline orientation of the \aalo\ substrate is crucial. For example, when growing \agox(0001)/\aalo(0001) after just 3-4 monolayers of pseudomorphically grown \aalgao(0001), a phase transformation into \bgao\ occurs, presumably due to strain relaxation of \aalgao\ above a film thickness ($d$) of $d > 2 \, \text{nm}$\cite{Schewski2015,Cheng2017,Karg2023}. The growth of \agao(01$\bar{1}$2) on $r$-plane \aalo(01$\bar{1}$2) results in $c$-plane facets forming from $d \approx$ 5 nm, which induces the nucleation and growth of \bgao\ on top of \aalo(01$\bar{1}$2) at $d \approx$ 217 nm \cite{Kracht2018}. Phase-pure \gox(11$\bar{2}$0) on $a$-plane \aalo(11$\bar{2}$0) by MBE has been limited to $d \approx 14 \, \text{nm}$\cite{Cheng2017}. The growth of \agao\ on $m$-plane \aal\ was demonstrated by MBE \cite{Jinno2021,McCandless2021} and MOCATAXY\cite{McCandless2023}. MBE works report stable \agao\ growth \cite{Jinno2021,McCandless2021}. Very recently, the growth of phase-pure \agao\ on \aal\ by MOCATAXY using In as the catalyst was shown to result in improved crystalline properties for \agao\ films, with (10$\bar{1}$1) facets formed on top of \agao(10$\bar{1}$0) \cite{McCandless2023}.

Following Ref.~\citenum{McCandless2023}, in this work we provide a comprehensive study on the kinetic and thermodynamic growth features of \agox\ and \aingao\ synthesized by MBE and MOCATAXY. The purpose of this work is to systematically investigate the growth parameter space accessible by MBE and MOCATAXY on the kinetic and thermodynamic growth processes that lead to \agao\ and \aingao\ formation. As a first approach, we study \agox\ films of $d \approx 50 \, \text{nm}$ formed on \aal. We find that both the oxygen-to-gallium flux ratio (\ogaflux) and the indium-to-gallium flux ratio (\ingafluxabb) determine the phase formation, the cation composition in \aingox\ and the surface features of the thin films grown. Single-crystalline \aga\ is achieved in metal-rich conditions and the presence of In, with step edges formed at the surface for higher growth temperatures (\tg) of 825°C. We develop a growth and phase diagram for the growth of phase-pure \agao\ and \aingao\ by MBE and MOCATAXY.

\section{Experimental}

\agao\ thin films were grown by MBE and MOCATAXY in a Riber Compact 12 system, equipped with an Oxford Applied Research HD25rf plasma source. Ga and In metals (6N purity) were supplied from standard effusion cells. The \aal\ substrates were backside sputter-coated with a Ti$_{0.1}$W$_{0.9}$ alloy of thickness $d$ $\approx$ 500 nm to enable radiative substrate heating during growth. All substrates were cleaned with deionized water and rinsed with isopropanol (IPA) to remove contamination from the dicing process, an ultrasonic bath in acetone for one minute, followed by an IPA rinse and dried by N$_2$. To eliminate residual surface contamination, a 10-minute plasma cleaning at 800°C, with O flux $\text{\fo} = 0.75$ standard cubic centimeters per minute (SCCM) and radio-frequency plasma-power $P_{\text{rf}} = 300 \,\text{W}$ was carried out \textit{in situ}. 

\tg\ was measured by a thermocouple located within the substrate heater and an optical pyrometer operating at a wavelength of $980 \, \text{nm}$. Reflection high-energy electron diffraction (RHEED) was used for \textit{in situ} growth monitoring and a retractable ion gauge located at the growth position to measure the Ga flux (\gaflux) and In flux (\influx) as beam equivalent pressure (BEP) in mbar. The O flux was supplied in SCCM, and active O (\oxflux) was generated by the radio frequency plasma source. To convert the measured BEP into physical units and to allow the reproduciblity of our results in other MBE systems, we convert \fga\ and \fin\ as $\text{mbar} \rightarrow \text{nm} \, \text{min}^{-1} \rightarrow \text{nm}^{-2} \text{s}^{-1}$ and \fo\ as $\text{SCCM} \rightarrow \text{nm} \, \text{min}^{-1} \rightarrow \text{nm}^{-2} \text{s}^{-1}$ (at given $P_{\text{rf}} = 300 \, \text{W}$) using the procedure established in Ref.~\citenum{Vogt2017thesis}. To achieve this, \gox\ and \ino\ calibration films were grown with conditions where all supplied cations and anions are incorporated into the respective thin film, i.e., when the sticking coefficients of Ga, In and O are unity. The density of the desired atoms in the unit cell is determined by crystallographic software (here VESTA \cite{Momma2011}). The particle fluxes of \fga, \fin\ and \fo\ can then be calculated as

\begin{equation}
	\phi_{i} = C \times \Gamma \times \rho_{i}
\end{equation}

with the growth rate ($\Gamma$) in nm min$^{-1}$, atomic density ($\rho_{i}$) of species $i = \text{Ga, In, O}$, and conversion factor $C$ = 1/60 to convert $\text{min} \rightarrow \text{sec}$. The maximum available active \fo\ for $\text{Ga} \rightarrow \text{\gox}$ and $\text{In} \rightarrow \text{\ino}$ oxidation can then be extracted from the $\Gamma$-peak at given \tg, i.e., \gr\ at stoichiometric growth conditions for \gox\ and In$_2$O$_3$\cite{Vogt2017thesis}. A summary of the calculated fluxes, as well as \ogaflux\ and \ingafluxabb, are given in Table \ref{tab:growthparams} and Table \ref{tab:samplenames}. 

\begin{table}
	\centering
	\caption{Summary of values for \fga, \fin, \fo\ and \tg\ for samples grown by MBE and MOCATAXY. For the MBE and MOCATAXY grown samples, the conversions from $\text{mbar} \rightarrow \text{nm}^{-2} \, \text{s}^{-1}$ are: $\phi$\textsubscript{Ga} = $5.2 \times 10 ^{-7}$ \, \text{mbar} \; $\mathrel{\widehat{=}} \, 4$ \fluxunit, $\phi_\text{{In}} \, = \, 1.2 \times 10 ^{-7} \, \text{mbar} \, \mathrel{\widehat{=}} \, 0.44$ \fluxunit, \oxflux\ = 0.25 SCCM $\,\mathrel{\widehat{=}} \, 1.6$ \fluxunit\ ($P$\textsubscript{rf} = 300 W). The active \oxflux\ in In-mediated catalysis is multiplied by 2.8 \cite{Vogt2017,McCandless2023}. The growth time was adjusted to control the layer thickness.}
	\label{tab:growthparams}
	\def\arraystretch{1.5}
	\begin{ruledtabular}
		\begin{tabular}{ccc}
    		\textbf{\footnotesize{Growth Parameters}}    & \textbf{\footnotesize{Conventional MBE}} & \textbf{\footnotesize{MOCATAXY}} \\ \hline
    		$\phi$\textsubscript{Ga} (\fluxunit)         & 4.0                                      & 4.0 \\
    		$\phi$\textsubscript{In} (\fluxunit)         & 0                                        & 0.4 -- 3.2 \\
    		$\phi$\textsubscript{O} (SCCM)               & 0.25 -- 0.75                             & 0.25 -- 0.75 \\
    		Active $\phi$\textsubscript{O} (\fluxunit)   & 1.6 -- 4.8                               & 4.5 -- 13.4 \\
    		$T$\textsubscript{G} (°C)                    & 775                                      & 775
		\end{tabular}
	\end{ruledtabular}
\end{table}
\begin{table}
	\centering
	\caption{Overview of samples studied in this work by MBE (\ingafluxabb\ = 0) and MOCATAXY (\ingafluxabb\ $>$ 0), with $R_{\text{In}}$ = \ingaflux, \ogaflux\ = \ogafluxdef\ and constant \gaflux\ = 4.0 \fluxunit. Samples A12, D2 and B12 are intermediate samples, and provide additional granularity to the results provided by primary samples where necessary.}
	\label{tab:samplenames}
	\def\arraystretch{1.5}
	\begin{ruledtabular}
		\begin{tabular}{c|cccc}
    		                            & \textbf{\ogaflux\ = 0.40} & \textbf{\ogaflux\ = 0.53} & \textbf{\ogaflux\ = 0.80} & \textbf{\ogaflux\ = 1.20} \\ \hline
    		\textbf{\ingafluxabb\ = 0.79}    & -                         & -                         & B4                        & - \\
    		\textbf{\ingafluxabb\ = 0.34}    & A3                        & -                         & B3                        & C3 \\
    		\textbf{\ingafluxabb\ = 0.11}    & A2                        & D2                        & B2                        & C2 \\
            \textbf{\ingafluxabb\ = 0.05}    & A12                       & -                         & B12                       & - \\
    		\textbf{\ingafluxabb\ = 0}       & A1                        & -                         & B1                        & C1
		\end{tabular}
	\end{ruledtabular}
\end{table}
High-resolution x-ray diffraction (HRXRD) and x-ray reflectometry were performed with a Philips X'Pert Pro-MRD using the Cu K$_{\upalpha 1}$ radiation to identify film thickness, crystal phase and determine the composition of \aingao. Surface morphologies were measured and root-mean squared (RMS) roughnesses determined by atomic force microscopy (AFM) in a Bruker Dimension Icon XR scanning probe microscope. An FEI Nova 200 focused ion beam was utilized to prepare selected samples for cross-sectional structural and chemical analysis. Scanning transmission electron microscopy (STEM) in high-angle annular dark-field imaging (HAADF) mode, using a probe-corrected Thermo Fisher Scientific Spectra 300 operating at an acceleration voltage of 300 kV, was employed to measure the atomic structure of the thin films, formed facets and interfaces. Spatially resolved energy-dispersive x-ray spectroscopy (EDX) was performed with the Super-X detection system, to measure the Al, Ga and In concentrations. Micro-Raman ($\upmu$-Raman) spectroscopy was performed for further phase identification and analysis using a Kimmon HeCd laser with a wavelength of 442 nm and a LabRAM HR Evolution confocal spectrometer.

\section{Results and Discussion}

\subsection{Growth Kinetics by MBE and MOCATAXY}

\begin{figure}
    \includegraphics[width=8.5cm]{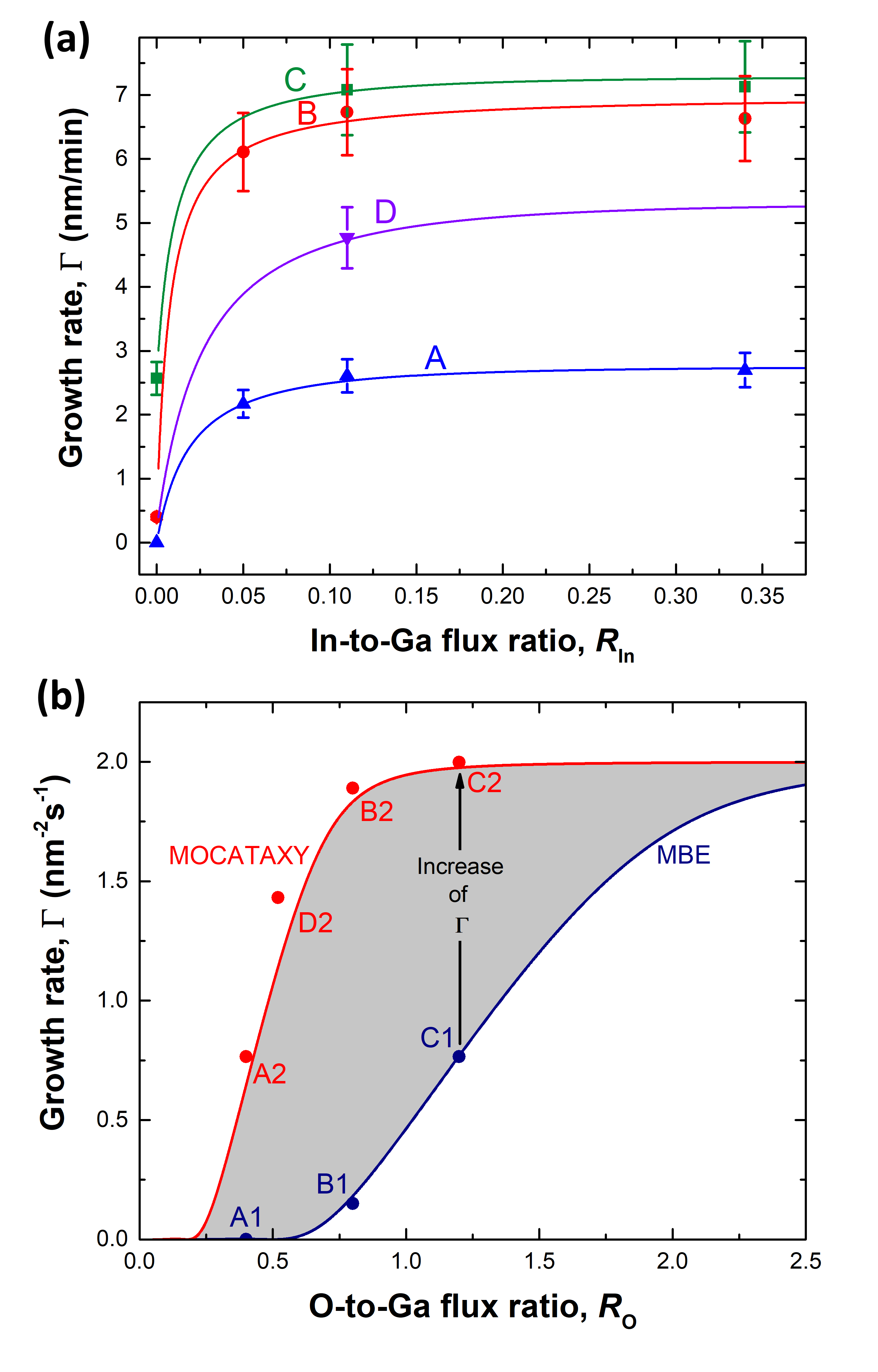}
    \caption{(a) \gr\ in nm min\textsuperscript{-1} as a function of \ingafluxabb\ at different \ogaflux. $\Gamma$ measured at \ingafluxabb\ = 0 corresponds to \gox\ growth by MBE and \ingafluxabb\ $>$ 0 to MOCATAXY. Lines are model calculations \cite{Vogt2017} and serve as a guide to the eye. (b) MBE and MOCATAXY models for \gr\ in \fluxunit\ as a function of \ogaflux. Models are shown for samples A1-C1 (MBE) and A2-C2 (MOCATAXY). The light-gray-shaded area depicts the expansion of the accessible growth window for \agao\ grown by MBE and MOCATAXY.}
    \label{fig:growthrate}
\end{figure}

In Fig. \ref{fig:growthrate}(a), \gr\ as a function of \ingafluxabb\ at different \ogaflux, at \tg\ = 775°C is plotted. See Table \ref{tab:samplenames} for the growth parameters of the displayed samples. All layers grown in this study have thicknesses $d \approx$ 50 nm, achieved by adjusting growth time after extracting \gr\ from calibration growths, and consulting the models shown in Fig. \ref{fig:growthrate}. For the MBE-grown samples, A1 (\ogaflux\ = 0.4), B1 (\ogaflux\ = 0.8) and C1 (\ogaflux\ = 1.2), at \ingafluxabb\ = 0, \gr\ increases with increasing \oxflux. At the growth conditions of A1, the nucleation and growth of \agox\ on \aaox(10$\bar{1}$0) is kinetically forbidden, as all active O is consumed to form the volatile suboxide Ga$_2$O which subsequently desorbs from the \aaox(10$\bar{1}$0) surface. With increasing \ogaflux, the formation of Ga$_2$O becomes less favored and \agox\ growth sets in for sample B1, with \gr\ further increasing for sample C1.

\gr\ as a function of \ogaflux\ is shown in Fig. \ref{fig:growthrate}(b). The MBE-grown samples (blue points represent experimental \gr, blue line represents modeled \gr) were grown under Ga-rich conditions, indicated by the increasing \gr\ with increasing \ogaflux. \gr\ plateaus in the O-rich regime, illustrated by the model at \ogaflux\ $>$ 1.5, and becomes limited by the supplied \gaflux. To expand the growth window of \agox, In is additionally supplied to the Ga-O growth system and MOCATAXY is employed. At the same \ogaflux\ and \tg, \gr\ increases with \ingafluxabb\ [indicated by the arrow in Fig. \ref{fig:growthrate}(b)] until \gr\ plateaus again, now limited by the supplied \fga\ and \fin.

Our experiment reveals that the available growth window of \aga/\aal\ is widened with larger \fo\ or \fin, see series A (\ogaflux\ = 0.4), B (\ogaflux\ = 0.8) and C (\ogaflux\ = 1.2) in Fig. \ref{fig:growthrate}(a). In Fig. \ref{fig:growthrate}(b), the light-gray-shaded area and arrow depict the expansion of the accessible growth window for \agao\ grown by MBE and MOCATAXY.

During MOCATAXY, the In as a catalyst provides more active O for metal oxidation to form the metal oxides, such as \agox\cite{McCandless2023}. Quantitatively, the available \fo\ for MOCATAXY-grown \gox\ can be 2.8 times larger than the \fo\ available for MBE-grown \gox\cite{McCandless2023,Vogt2017,Vogt2022}. We note that the models shown in Fig. \ref{fig:growthrate} use arbitrary kinetic parameters, similar to the models shown in Ref. \citenum{McCandless2023}. The model closely follows the experimental \gr\ values as a function of \ogaflux\ for the MBE and MOCATAXY samples. We find a very good correspondence to the experimental data when assuming full Ga incorporation; 2 atoms \fluxunit\ for sample C2 by MOCATAXY, with the higher oxygen flux. It must be noted that the contribution of a 7 met.\% incorporation of In in this sample, as further discussed below, is considered in the model.

\subsection{Surface Morphology}

\begin{figure*}
	\includegraphics[width=17cm]{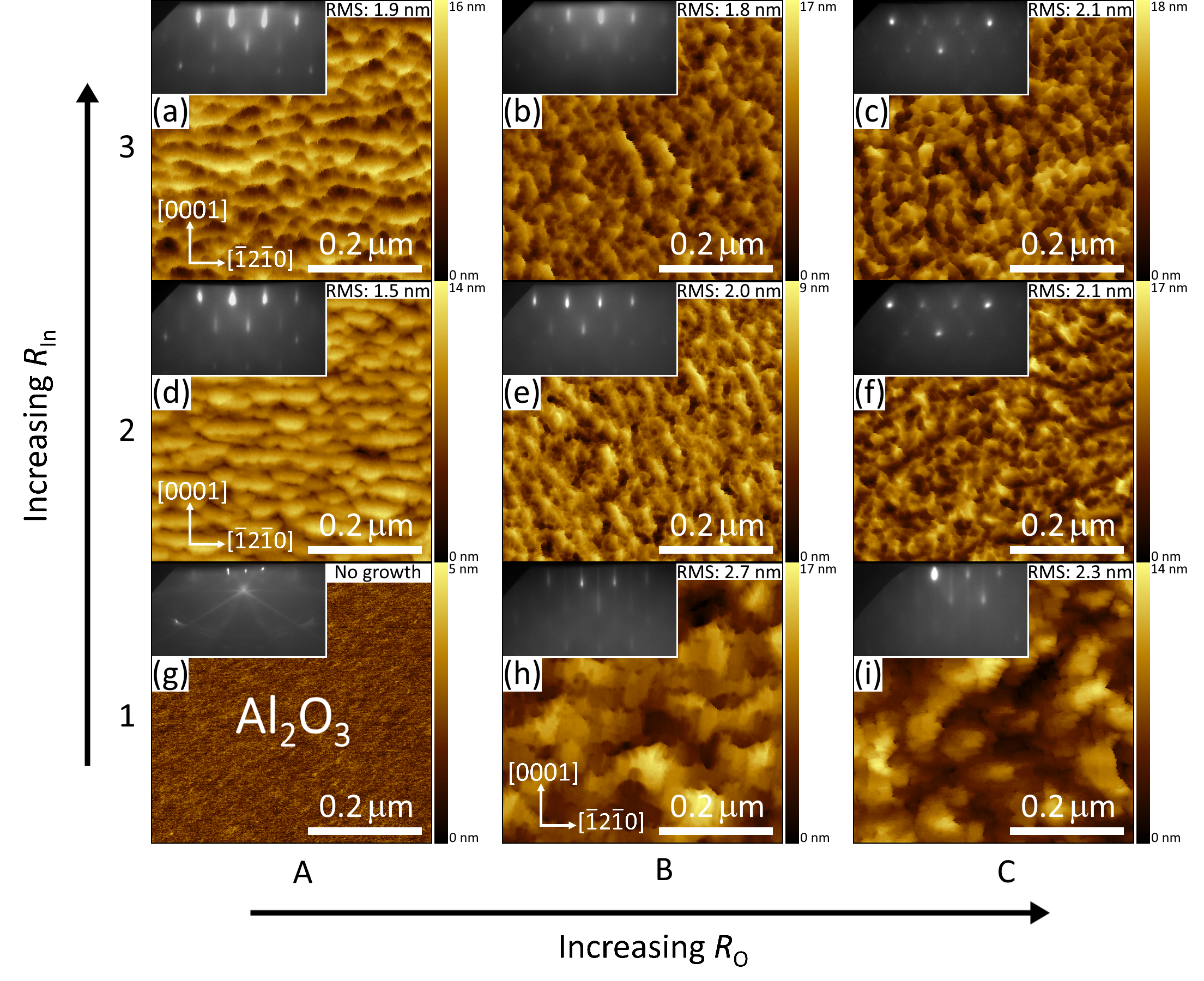}
	\caption{0.5 $\upmu$m $\times$ 0.5 $\upmu$m AFM images of samples A1-C3. Insets show corresponding RHEED images, taken along the [11$\bar{2}$0] azimuth. RMS surface roughness values correspond to the 5 $\upmu$m $\times$ 5 $\upmu$m AFM images shown in the supplementary material, Fig. S1.}
	\label{fig:afmscans}
\end{figure*}

The impact of \ingafluxabb\ and \ogaflux\ on the surface morphology of the same samples as in Fig. \ref{fig:growthrate} (except A12, B12 and D2) is depicted in Fig. \ref{fig:afmscans}. RMS roughness is determined by AFM over a \linebreak 5 $\upmu$m $\times$ 5 $\upmu$m range, shown in the supplementary material, Fig. S1. Corresponding RHEED patterns of all the samples are shown in the insets of Fig. \ref{fig:afmscans}. Both samples C2 and C3 exhibit additional weak RHEED spots between the main \agox-related spots [insets in Fig. \ref{fig:afmscans}(f) and \ref{fig:afmscans}(c)]. This indicates \bgox\ formation, as discussed in the following sections. Rough surfaces (RMS $\geq$ 2.3 nm) are observed in MBE samples B1 and C1, whereas the surface RMS is improved in the MOCATAXY samples. Samples A2, A3 and D2 (see Fig. S2(a) in the supplementary material) exhibit further improved surface morphologies with lower RMS roughnesses (RMS $\leq$ 1.9 nm) and extended surface facets. The extended features are oriented parallel to the [$\bar{1}$2$\bar{1}$0] (a-plane) direction of \agox, indicating that In has a pronounced surfactant effect along the [$\bar{1}$2$\bar{1}$0] direction with respect to the [0001] direction. The role of In in surface roughening/smoothing (i.e. as a (anti)surfactant) was previously reported during MOCATAXY growth of \gox\ or \algox \cite{Vogt2018,McCandless2023,Karg202309}. In acting as a surfactant in the MOCATAXY regime for \ogaflux\ $\leq$ 0.53 may be attributed to an enhanced adatom mobility when In is supplied, similar to the well-known In-Ga-N system \cite{Neugebauer2003}.


\subsection{Crystalline Phase and In Incorporation Into \agox}

\begin{figure}
	\includegraphics[width=8.5cm]{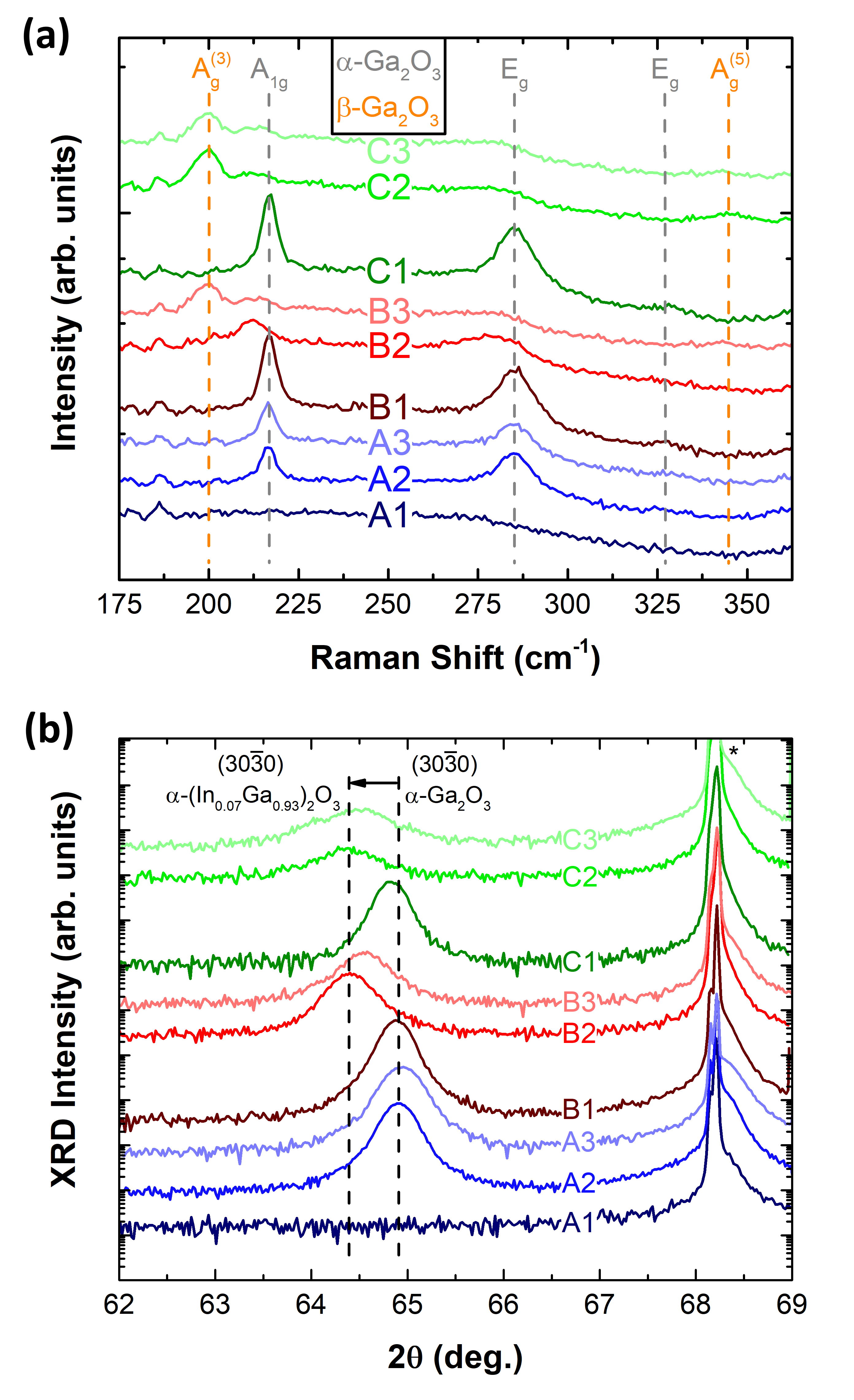}
	\caption{(a) $\upmu$-Raman spectra acquired for samples A1-C3, with \agox\ and \bgox\ Raman modes marked by gray and orange dashed lines respectively. (b) Symmetric \xrdqw\ HRXRD scans recorded for samples A1-C3. The peak of the \aaox\ substrate is marked by an asterisk. The \aingao\ peaks are marked by dashed lines, for $x$ = 0.07 and $x$ = 0.}
	\label{fig:symmetricxrdandraman}
\end{figure}

In order to investigate the crystalline phase of the samples in Fig. \ref{fig:afmscans}, $\upmu$-Raman spectroscopy was employed, as shown in Fig. \ref{fig:symmetricxrdandraman}(a). Sample A1 exhibits no Raman modes beyond those of the \aaox\ substrate \cite{Palanza2008} (at higher Raman shifts than the displayed range), because no growth has occurred. Samples A2, A3, B1 and C1 all exhibit additional intense and well-defined Raman peaks at 217 cm$^{-1}$, 285 cm$^{-1}$ and 327 cm$^{-1}$, which are characteristic modes of the corundum \agox\ structure \cite{Cusco2015}. In samples B2, B3, C2 and C3, the peak positions of the corundum Raman modes present much lower relative intensities, higher full widths at half maximum (FWHM, e.g. 11 cm$^{-1}$ for the 217 cm$^{-1}$ peak in sample B2, compared to 5 cm$^{-1}$ in lower \oxflux\ and \influx\ samples) and Raman redshifts, which can be explained by the substantial amount of In in these \agox\ films, in agreement with In concentrations extracted from HRXRD discussed below. In addition, all of these samples except B2 display intense peaks at Raman shifts of 199 cm$^{-1}$-200 cm$^{-1}$ and 344 cm$^{-1}$, that are assigned to the A$_{\text{g}}^\text{{(3)}}$ and A$_{\text{g}}^\text{{(5)}}$ modes of monoclinic $\upbeta$-(In)\gox, respectively \cite{Kumar2014,Kranert2016}. This reveals the presence of $\upbeta$-\ingoxcon\ in the MOCATAXY samples when high \ingafluxabb\ and \ogaflux\ are provided. It should be noted that trial growths with lower thicknesses result in phase-pure $\upalpha$-\ingoxcon\ layers which implies that $\upbeta$-\ingoxcon\ heteroepitaxially grows on $\upalpha$-\ingoxcon\ after a critical thickness. This is investigated further in the following section.

Figure \ref{fig:symmetricxrdandraman}(b), shows symmetric \xrdqw\ HRXRD scans. No (30$\bar{3}$0) \agox\ reflex is detectable in sample A1, as the nucleation and growth of \gox\ is kinetically forbidden. Samples grown by MBE with larger \oxflux\ (B1, C1) exhibit the expected (30$\bar{3}$0) \agox\ reflex at 2$\uptheta$ = 64.9° \cite{McCandless2023,Jinno2021}. No other crystalline phases of \gox\ were detected. However, MOCATAXY samples grown with \ingafluxabb\ $\geq$ 0.11 and \ogaflux\ $\geq$ 0.80 (samples B2 and B3, C2 and C3) show a shift in the diffraction angles to 22$\uptheta$ = 64.4°-64.6° due to In incorporation in the \agox\ layer, while those with \ogaflux\ $\leq$ 0.53 (samples A2, D2 and A3) still exhibit the (30$\bar{3}$0) reflex at 64.9° (see the blue spectrum, and Fig. S2(b)). 

For fully relaxed films, Vegard's law for the a-lattice parameter may be used to extract In concentrations from diffraction peak positions. From HRXRD, the average In concentration in the \aingao\ samples is estimated to $x \approx 0.07$ for sample B2. No strain is considered for this estimation, as a reciprocal space map (RSM), shown in Fig. S3 in the supplementary material, shows the layer to be fully relaxed. An independent concentration determination using STEM-EDX yielded $x$ = 0.07 $\pm$ 0.01 (see Fig. S6b), which is in good agreement with this result. Once \influx\ is further increased at constant \ogaflux\, a gradual shift to higher 2$\uptheta$ angles, up to 64.7° in samples B2-B4, is observed (see Fig. S4), i.e. a lower In incorporation is measured, down to $x \approx 0.03$, despite a higher \ingafluxabb\ supplied. We attribute this behavior to the In solubility limit in \agox\ being reached for our growth conditions, and excess In forming the suboxide In$_{\text{2}}$O that may desorb from the growth surface at these metal-rich flux conditions. A recent investigation in $\upepsilon$-\ingoxcon\ found analogous behavior at high \influx, with In concentrations approaching $x \approx$ 0 \cite{Karg202309}. The maximum In concentration of $x \approx 0.08$ in sample B2 agrees with the reported maximum In concentration for mist CVD-grown phase-pure $\upalpha$-\ingoxcon\ \cite{SUZUKI2014670}. In contrast, MOCATAXY-grown samples A2, A3 and D2 (blue spectra in Fig. \ref{fig:symmetricxrdandraman}, and Fig. S2(b) in the supplementary material) are phase-pure \agox\ and have lower surface RMS roughnesses. This effect can be correlated to the thermodynamics of the In incorporation in $\upbeta$-\ingoxcon, reported in Ref.~\citenum{Vogt2016}. There, it was found that a low metal-to-oxygen flux ratio (\rme) leads to $\upbeta$-\ingoxcon, while \bgox\ was obtained when \rme\ = (\influx\ + \gaflux)/\oxflux\ = 2. This is also the case here; sample D2 was grown with \rme\ = 2.1 and no observable In incorporation while sample B2 (red spectrum in Fig. \ref{fig:symmetricxrdandraman}), with \rme\ = 1.4, contains $\sim$7\% In.

Rocking curves for representative samples can be found in the supplementary material, Fig. S5. The MBE-grown \agox\ (B1 and C1) and the MOCATAXY-grown \agox\ (A2) samples exhibit the same FWHM, suggesting MOCATAXY does not provide a measurable improvement in the mosaicity at these growth conditions. In contrast, $\upalpha$-\ingoxcon\ samples [e.g., B2 in Fig. S5(c)] show markedly increased FWHM values, due to varying In concentrations within layers and greater mosaic spreads.

The AFM and HRXRD data suggest \rme\ $\geq$ 2 is required to take full advantage of In being a catalyst and surfactant for the growth of \agox. Conditions with \rme\ $<$ 2 result in the sub-optimal role of In as a surfactant with In incorporation in the layer up to its solubility limit, while too high \rme\ results in\ no growth. These \rme-dependent behaviors closely follow those observed for \bgox \cite{Vogt2016}.

\subsection{Faceting and Interfaces}

\begin{figure}
	\includegraphics[width=8.5cm]{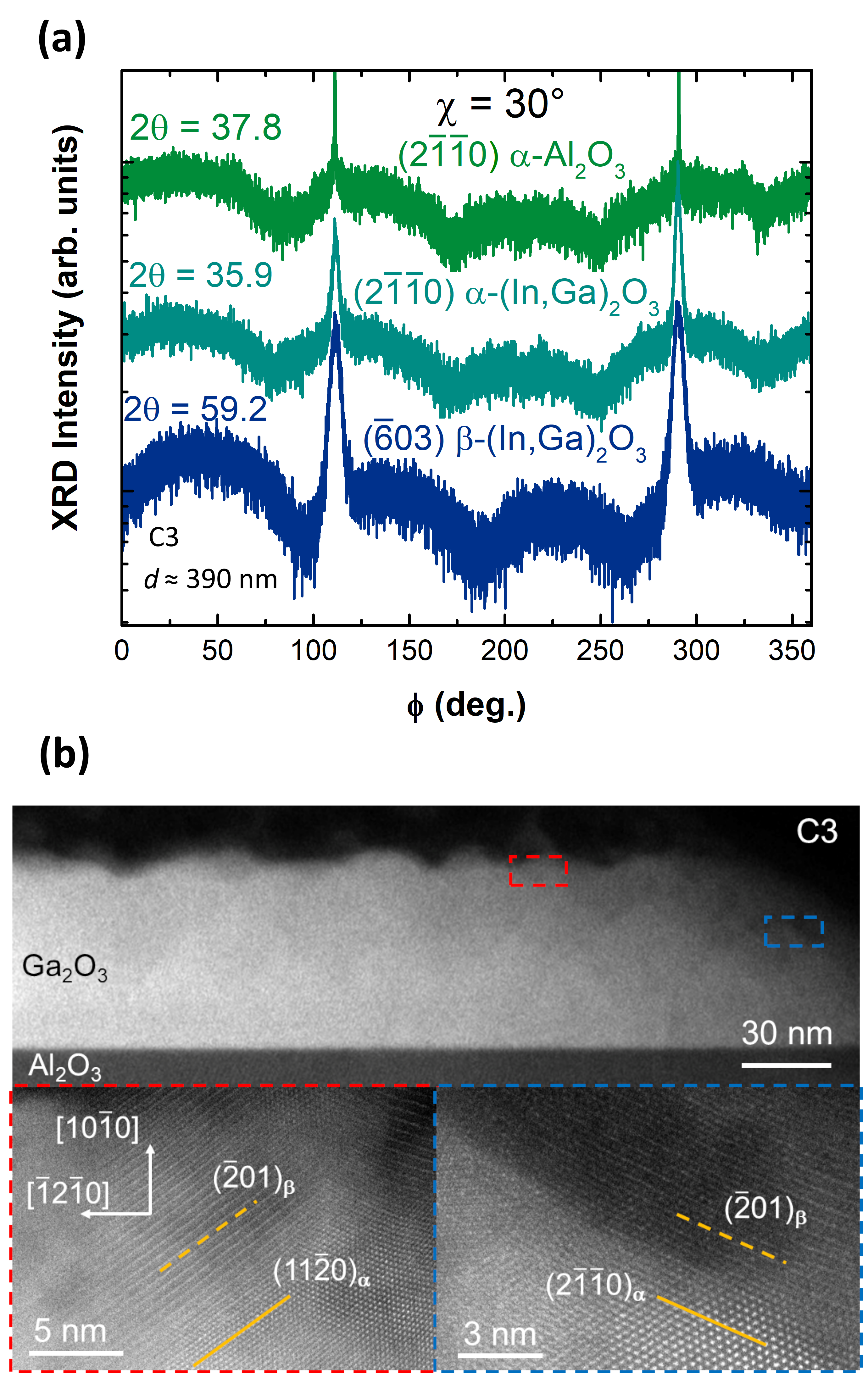}
	\caption{Sample C3: (a) For a film with $d \approx$ 390 nm, asymmetric HRXRD $\phi$ scans ($\chi$ = 30°) of the (2$\bar{1}\bar{1}$0) peak of the \aaox\ substrate, the (2$\bar{1}\bar{1}$0) peak of $\upalpha$-\ingoxcon\ and the ($\bar{6}$03) peaks of $\upbeta$-\ingoxcon. (b) Cross-sectional HAADF overview and magnified images, showing two examples of the a-plane faceting and its epitaxial relation to ($\bar{2}$01) $\upbeta$-\ingoxcon.}
	\label{fig:C3}
\end{figure}

To identify the mechanisms leading to $\upbeta$-\ingoxcon\ heteroepitaxial growth or phase-pure \agox\ on \aaox(10-10), asymmetric HRXRD scans and STEM analysis were performed. In Fig. \ref{fig:C3}(a), asymmetric HRXRD $\phi$-scans at $\chi$ = 30° (i.e. parallel to the a-plane of \aaox) for a sample equivalent to C3, but with $d$ = 390 nm to maximize \bgox\ growth and signal, are shown. The diffraction peaks of a-plane (2$\bar{1}\bar{1}$0) \aaox\ and $\upalpha$-\ingoxcon\ are measured since the (2$\bar{1}\bar{1}$0) plane forms 30° with respect to the growth plane (10$\bar{1}$0). Additionally, the ($\bar{6}$03) $\upbeta$-\ingoxcon\ reflex at 2$\theta$ = 59.2° is clearly visible in this orientation. Hence, the \aaox\ substrate, the initial $\upalpha$-\ingoxcon\ film and the subsequent $\upbeta$-\ingoxcon\ layer are identified with the epitaxial relationship ($\bar{2}$01) $\upbeta$-\ingoxcon\ $\parallel$ (2$\bar{1}\bar{1}$0) $\upalpha$-\ingoxcon. Both (2$\bar{1}\bar{1}$0) planes in Fig. \ref{fig:C3}(a) show two-fold symmetry in the 360° $\phi$ scan and appear at the same $\phi$ angle, which implies that there are no rotational domains in the \aingox\ layer (nor in the \agox\ samples) on m-plane \aaox \cite{Grundmann2010}.

Figure \ref{fig:C3}(b) shows cross-sectional HAADF overview and magnified images of sample C3. Two distinct phases are identified: $\upalpha$-\ingoxcon\ [see below solid orange lines in Fig. \ref{fig:C3}(b)] and $\upbeta$-\ingoxcon\ [see above dashed orange lines in Fig. \ref{fig:C3}(b)]. The (2$\bar{1}\bar{1}$0) facet of the $\upalpha$-phase is identified as the growth surface for $\upbeta$-\ingoxcon. In the HAADF image, we could only resolve the atomic distances between ($\bar{2}$01) planes of $\upbeta$-\ingoxcon, but not perpendicular distances. We expect that these perpendicular distances should be resolvable in [010] and [132]-type orientations. Therefore, the most likely epitaxial relation between $\upalpha$-\ingoxcon\ and  $\upbeta$-\ingoxcon\ is ($\bar{2}$01) $\upbeta$-\ingoxcon\ $\parallel$ (2$\bar{1}\bar{1}$0) $\upalpha$-\ingoxcon\ and  [102] $\upbeta$-\ingoxcon\ $\parallel$ [0001] $\upalpha$-\ingoxcon. Hence, at a film thickness $d$ $\approx$ 50 nm, nucleation of the $\upbeta$-phase only occurs in the MOCATAXY region, when there is a sufficiently large \ogaflux\ and \ingafluxabb\ that allows for In incorporation. STEM-EDX is shown for sample C3 in the supplementary material, Fig. S6(a). Delayed In incorporation is observed in the film. The maximum concentration of In measured in C3, $x$ $\approx$ 0.12, occurs after $d \approx$ 25 nm. Accordingly, this higher In incorporation occurs in \bgox. Higher metal concentrations of In in $\upbeta$-\ingoxcon\ were previously theoretically predicted \cite{Peelaers2015} and experimentally achieved by MBE \cite{Kranert2014}.

Cross-sectional HAADF images and STEM-EDX for sample B2 is shown in the supplementary material, Fig. S6(b) and Fig. S7. (2$\bar{1}\bar{1}$0) a-plane facets are also identified in this sample, without the growth of $\upbeta$-\ingoxcon, suggesting that the critical $\upalpha$-\ingoxcon\ thickness before \bgox\ nucleation is about to be reached for these growth conditions. Such critical thickness, $d \approx$  50 nm, is double that of e.g. sample C3, which implies that the a-plane facet formation and subsequent $\upbeta$-\ingoxcon\ growth is delayed when reducing \ingafluxabb\ to 0.11 and \ogaflux\ to 0.8.


\begin{figure}
	\includegraphics[width=8.5cm]{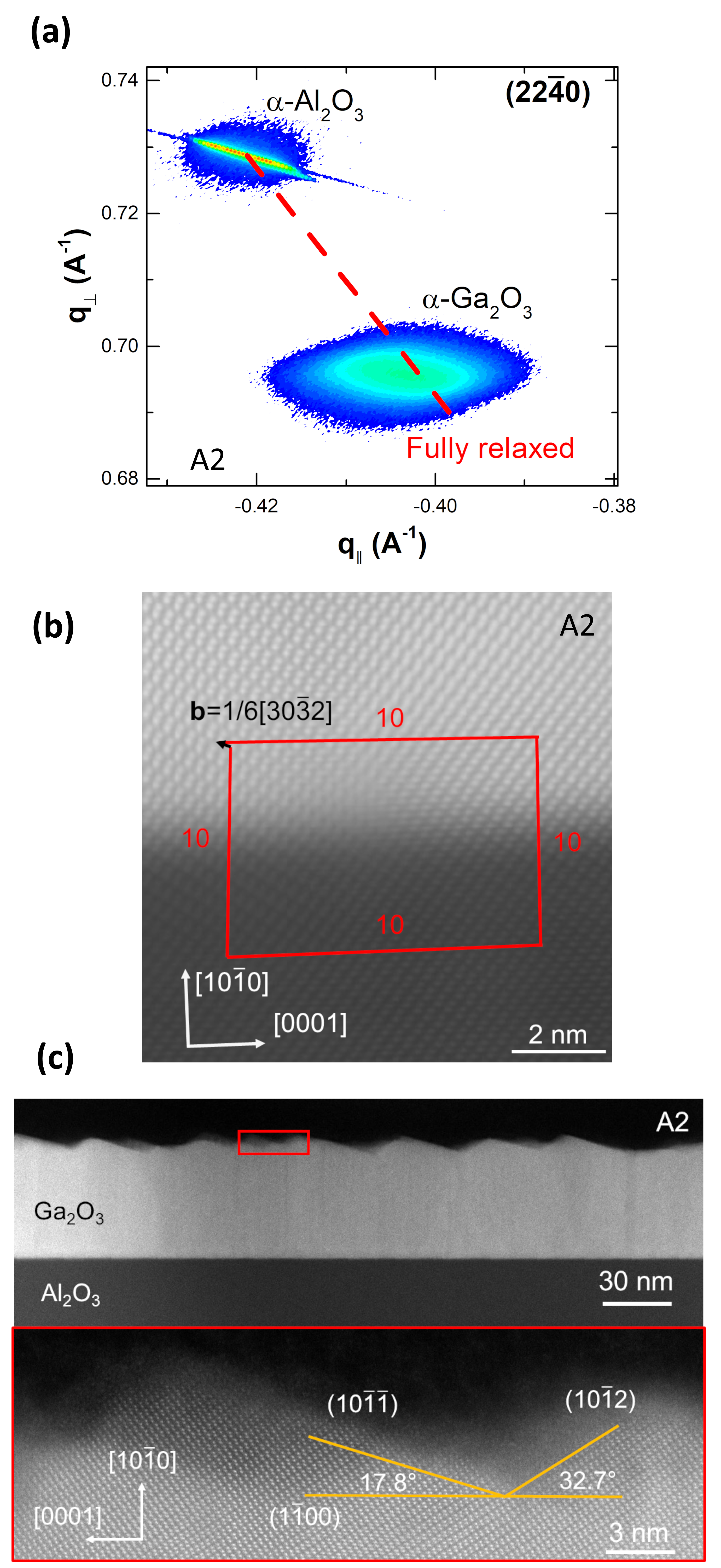}
	\caption{Sample A2: (a) RSM (q$_{\parallel}$, q$_{\perp}$) around the (22$\bar{4}$0) reflex of sample A2, displaying full relaxation of the \agox\ film. (b) Magnified cross-sectional HAADF image of the substrate-layer interface, with the Burgers vector $\mathbf{b}$ noted. (c) Cross-sectional HAADF overview and magnified images, showing the formation of (10$\bar{1}\bar{1}$) and (11$\bar{1}$2) facets, indicated in yellow.}
	\label{fig:A2}
\end{figure}

An HRXRD RSM of sample A2 around the (22$\bar{4}$0) \agox\ reflex is shown in Fig. \ref{fig:A2}(a). The dashed line intersects the fully-relaxed \aaox\ and \agox\ reciprocal lattice points. The results indicate that the 50 nm \agox\ layer is fully relaxed. Figure \ref{fig:A2}(b) shows the substrate-film interface in sample A2, with the presence of a misfit dislocation. The film relaxes at the interface to reduce the elastic strain energy between \gox(10$\bar{1}$0) and \aox(10$\bar{1}$0) \cite{Hull1992}. The magnitude of the dislocation shown in Fig. \ref{fig:A2}(b) can be represented by the Burgers vector \textbf{b} = 1/6[30$\bar{3}$2].

Figure \ref{fig:A2}(c) shows the cross-sectional HAADF overview and magnified images at the surface of sample A2. Single-crystalline material is identified in the observed regions over the range of a few microns. Unlike sample C3, no a-plane facets or secondary phase formation are present in A2. Instead, (10$\bar{1}\bar{1}$) facets are observed on the surface, reflecting the morphology observed by AFM. The equivalency of \gox(10$\bar{1}$1) and \gox(10$\bar{1}\bar{1}$), due to symmetry, confirms the noted surface facet orientation we previously reported \cite{McCandless2023}. STEM-EDX cross-sectional maps and line scans for sample A2 are shown in the supplementary material, Fig. S6(c). In accordance with the HRXRD and $\upmu$-Raman analysis, In does not substantially incorporate in the film, i.e., below 1 at. \%.

\subsection{MBE and MOCATAXY Phase Diagram of \gox}

\begin{figure}
    \includegraphics[width=8.5cm]{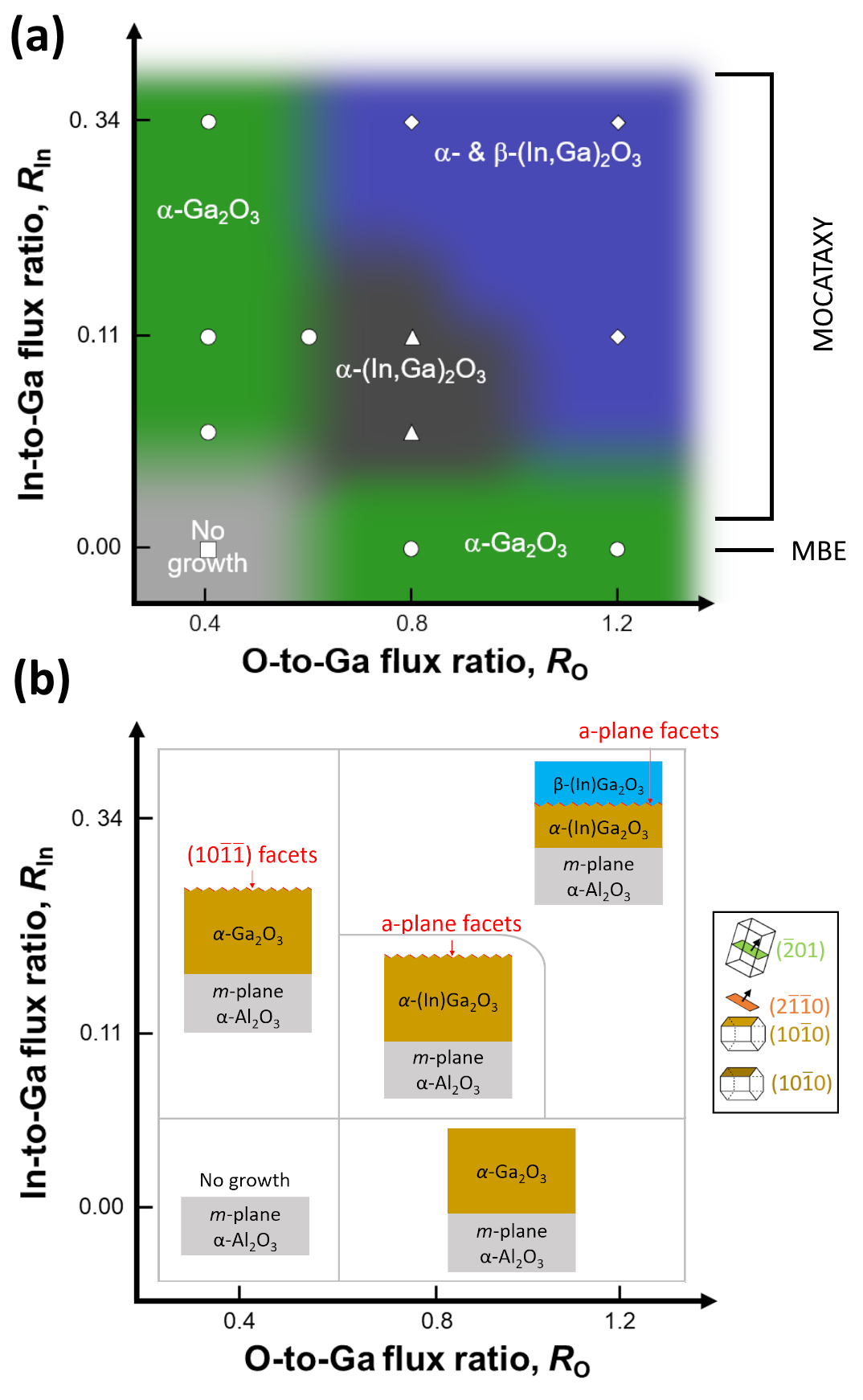}
 	\caption{(a) Measured and developed growth and phase diagram of $\upalpha$-\ingoxcon\ with $x \leq$ 0.08 at $T_G$ = 775°C, \gaflux\ = 4.0 \fluxunit\ and $d \approx$ 50 nm, projected onto the 2-dimensional parameter space spanned by the In-to-Ga flux ratio (\ingafluxabb) and O-to-Ga flux ratio (\ogaflux). Four major regimes are indicated: (i) no growth, (ii) single-crystalline \agox, (iii) phase-pure $\upalpha$-\ingoxcon\ and (iv) $\upalpha$-\ingoxcon\ and $\upbeta$-\ingoxcon\ growth regimes. (b) Schematic of \gox\ phase growth and faceting on m-plane \aaox, for different \ingafluxabb\ and \ogaflux, and $d$ = 50 nm. The epitaxial relationships are shown by an illustration on the right side of the figure. Light gray lines serve as a guide to the eye from Fig. \ref{fig:growthdiagram}(a).}
 	\label{fig:growthdiagram}
\end{figure}

To guide the growth of \agox(10$\bar{1}$0)/\aaox(10$\bar{1}$0), Fig. \ref{fig:growthdiagram}(a) shows a phase diagram that encompasses the $d$ $\approx$ 50 nm films studied in this work (Table \ref{tab:samplenames}). Up to this thickness, only MOCATAXY-grown samples with high \ingafluxabb\ \textit{and} high \ogaflux\ exhibit heteroepitaxial $\upbeta$-\ingoxcon\ growth. At lower \ingafluxabb\ = 0.05 and \ogaflux\ = 0.80 [see Fig. S2(a)], In also incorporates into the grown layers but no secondary $\upbeta$-phase is observed. In that sample, the solubility limit of In ($x \approx$ 0.08) in \agox\ is already reached. For \ogaflux\ $\leq$ 0.53, independent of \ingafluxabb, phase-pure \agox\ layers with In incorporation below 1 at.\% are obtained. 

\begin{figure}[h!]
    \includegraphics[width=8.5cm]{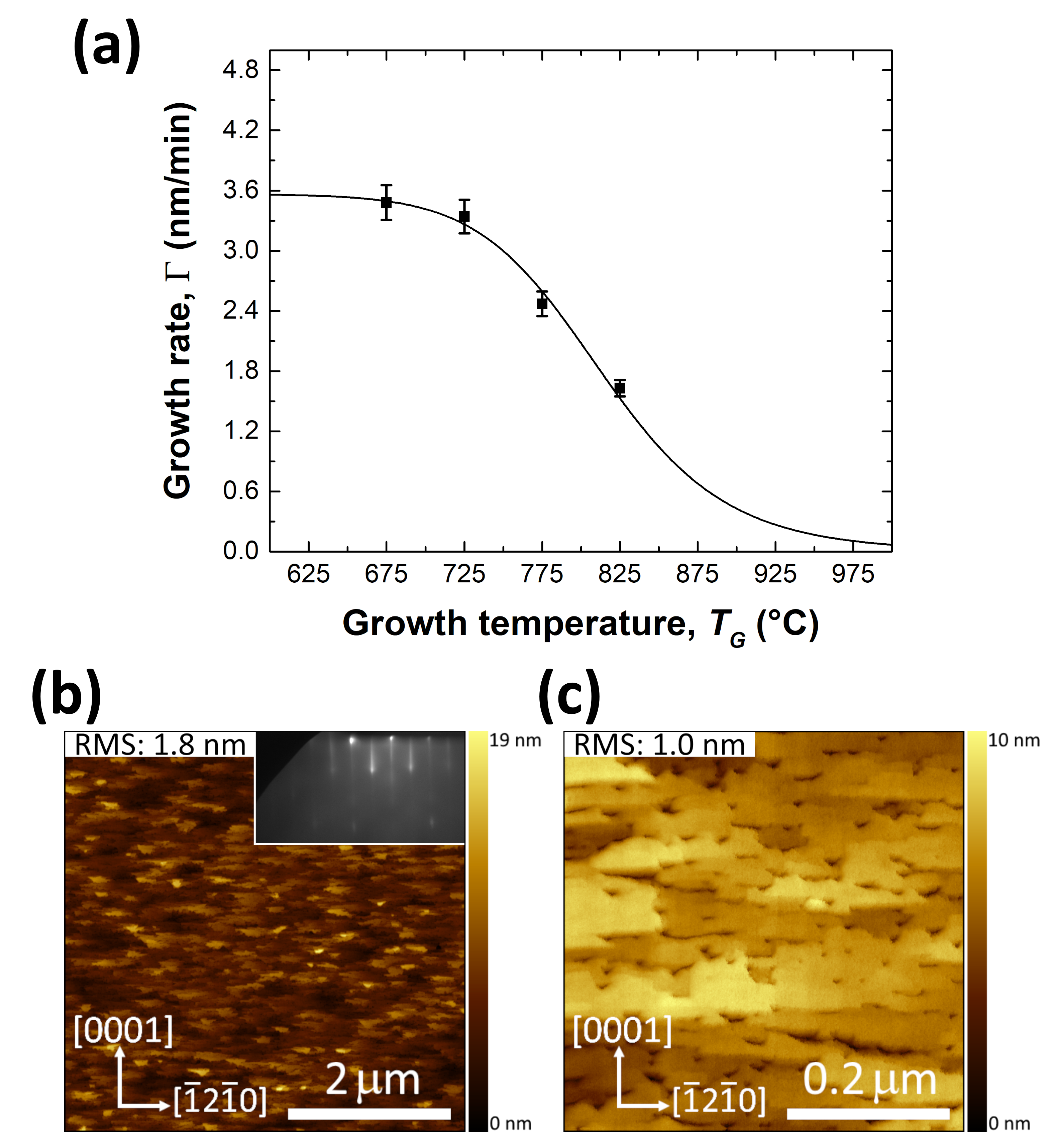}
 	\caption{(a) Temperature-dependent growth rate for samples grown with parameters of A2, line fit is a model fit and serves as a guide to the eye. (b) 5 $\upmu$m $\times$ 5 $\upmu$m AFM image of A2 grown at 825°C, inset shows RHEED pattern taken along [11$\bar{2}$0] azimuth, and (c) 500 nm $\times$ 500 nm AFM image of the same sample, exhibiting well-defined terraces and step edges.}
    \label{fig:tdep}
\end{figure}

Figure \ref{fig:growthdiagram}(b) sketches the growth-parameter-dependent \agox\ surface faceting and $\upalpha$-\ingoxcon-to-$\upbeta$-\ingoxcon\ phase transition critical thickness for our $d$ = 50 nm samples. It is possible to achieve single-crystalline \agox\ both by MBE and MOCATAXY For the latter it is necessary to supply a relatively low oxygen flux, \ogaflux\ $\leq$ 0.53. Under such conditions, (10$\bar{1}\bar{1}$)-plane facets appear on the \agox\ surface, similar to what was reported in Ref.~\citenum{McCandless2023}. It must be noted that, similar to  a recent PLD study on \agao \cite{Petersen2023}, we have detected inhomogeneous \bgoxmitin\ formation on thicker ($>$ 100 nm) samples under these growth conditions, which is an open topic beyond the scope of this work. Our results assert the need to tune the growth parameters, aiming for a smooth layer-by-layer growth that prevents the formation of facets where the secondary phase growth takes place. This is partially achieved in the A2 sample at \tg\ = 825°C, as presented in the next section.

\subsection{Temperature-Dependent Growth Series}

To investigate the effect of \tg\ on our optimized phase-pure \agox\ film (sample A2), a \tg-series at such optimal \gaflux, \influx\ and \oxflux\ is performed. Figure \ref{fig:tdep}(a) shows a decrease in \gr\ with increasing \tg\ due to increased Ga$_2$O desorption \cite{Vogt2018model}. The rocking curve for the sample grown at \tg\ = 825°C is shown in the supplementary material, Fig. S5(e), indicating an improvement in the layer's mosaicity with respect to the equivalent sample at \tg\ = 775°C, Fig. S5(a). Figure \ref{fig:tdep}(b) and \ref{fig:tdep}(c) show 5 $\upmu$m $\times$ 5 $\upmu$m and 0.5 $\upmu$m $\times$ 0.5 $\upmu$m AFM images of the sample grown at 825°C. Its surface exhibits well-defined terraces and step edges, also observable by the modulated streaky RHEED pattern. The step size can vary by a significant amount across the surface, with most falling in the range 3-5 nm, corresponding to approximately 14-23 atomic planes. Again, revealed by AFM and HRXRD, these features are aligned parallel to the [$\bar{1}$2$\bar{1}$0] direction [marked in Fig. \ref{fig:tdep}(b) and \ref{fig:tdep}(c)]. Although the surface RMS roughness of this sample is higher than the value for sample A2 (same conditions but $T_G$ = 775°C), certain areas such as the one shown in Fig. \ref{fig:tdep}(c) possess markedly lower RMS values of 1.0 - 1.3 nm, pointing towards the possibility of achieving smoother \agox\ thin films at this \tg. The drastic \tg-driven surface morphology change with respect to sample A2 can be attributed to enhanced adatom surface mobility.

As shown in the symmetric HRXRD spectrum of the (30$\bar{3}$0) reflex of these \tg-dependent samples, in Fig. S8, no additional phases and no \agox\ peak shifts above the experimental uncertainty were present, which indicates the dominant role of the In-Ga-O kinetics with respect to thermodynamics in the formation of phase-pure \agox.

\section{Conclusions}

The effect of varying the In and O fluxes on the formation of \agox\ on m-plane \aaox\ by MBE has been systematically investigated. Using MOCATAXY, three growth regimes are identified, resulting in: (i) single-crystalline \agox, (ii) phase-pure $\upalpha$-\ingoxcon, with up to $x$ = 0.07 $\pm$ 0.01, and (iii) $\upalpha$-\ingoxcon\ + $\upbeta$-\ingoxcon\ growth on a-plane $\upalpha$-\ingoxcon\ facets. To grow single-crystalline \agox\ by In-assisted MOCATAXY and avoid considerable In incorporation, \ogaflux\ $\leq$ 0.53 is necessary. Under these conditions, In acts as a surfactant along the [$\bar{1}$2$\bar{1}$0] direction. Higher \tg\ result in step-like growth, with the potential of achieving smoother \agox\ samples that prevent secondary phase formation in thicker films. Such understanding and optimization of the growth kinetics and thermodynamics of \agox\ on \aaox\ is necessary in order to realize high quality films and heterostructures based on this ultra-wide band gap material system. 

\section*{Acknowledgments}
M. A.-O. acknowledges financial support from the Central Research Development Fund (CRDF) of the University of Bremen.

J. P. M. acknowledges the support of a National Science Foundation Graduate Research Fellowship under Grant No. DGE–2139899

\section*{Conflict of Interest Statement}

The authors have no conflicts to disclose.

\section*{Data Availability Statement}

The data that supports the findings of this study are available from the corresponding author upon reasonable request.

\bibliography{references}

\end{document}


\preprint{AIP/123-QED}

\title{Growth, catalysis and faceting of \agao\ and \aingao\ on $m$-plane \aalo\ by molecular beam epitaxy}

\author{Martin S. Williams}
\affiliation{ 
Institute of Solid State Physics, University of Bremen, Otto-Hahn-Allee 1, 28359, Bremen, Germany
}%

\author{Manuel Alonso-Orts}
\affiliation{ 
Institute of Solid State Physics, University of Bremen, Otto-Hahn-Allee 1, 28359, Bremen, Germany
}%
\affiliation{MAPEX Center for Materials and Processes, University of Bremen, Bibliothekstraße 1, 28359, Bremen,
Germany}

\author{Marco Schowalter}
\affiliation{ 
Institute of Solid State Physics, University of Bremen, Otto-Hahn-Allee 1, 28359, Bremen, Germany
}%

\author{Alexander Karg}
\affiliation{ 
Institute of Solid State Physics, University of Bremen, Otto-Hahn-Allee 1, 28359, Bremen, Germany
}%

\author{Sushma Raghuvansy}
\affiliation{ 
Institute of Solid State Physics, University of Bremen, Otto-Hahn-Allee 1, 28359, Bremen, Germany
}%

\author{Jon P. McCandless}
\affiliation{ 
School of Electrical and Computer Engineering, Cornell University, 229 Phillip's Hall, 14853, New York, United States of America
}%

\author{Debdeep Jena}
\affiliation{ 
School of Electrical and Computer Engineering, Cornell University, 229 Phillip's Hall, 14853, New York, United States of America
}%
\affiliation{ 
Department of Material Science and Engineering, Cornell University, Bard Hall, 14853, New York, United States of America
}%
\affiliation{
Kavli Institute at Cornell for Nanoscale Science, Cornell University, Ithaca, New York, NY-14853, United States of America
}%

\author{Andreas Rosenauer}

\author{Martin Eickhoff}
\affiliation{ 
Institute of Solid State Physics, University of Bremen, Otto-Hahn-Allee 1, 28359, Bremen, Germany
}%
\affiliation{MAPEX Center for Materials and Processes, University of Bremen, Bibliothekstraße 1, 28359, Bremen,
Germany}

\author{Patrick Vogt}
\affiliation{ 
Institute of Solid State Physics, University of Bremen, Otto-Hahn-Allee 1, 28359, Bremen, Germany
}%

\date{\today}

\maketitle

\centering
\renewcommand{\thefigure}{S1}

\captionsetup{type=figure}
\includegraphics[width=\textwidth,page=1]{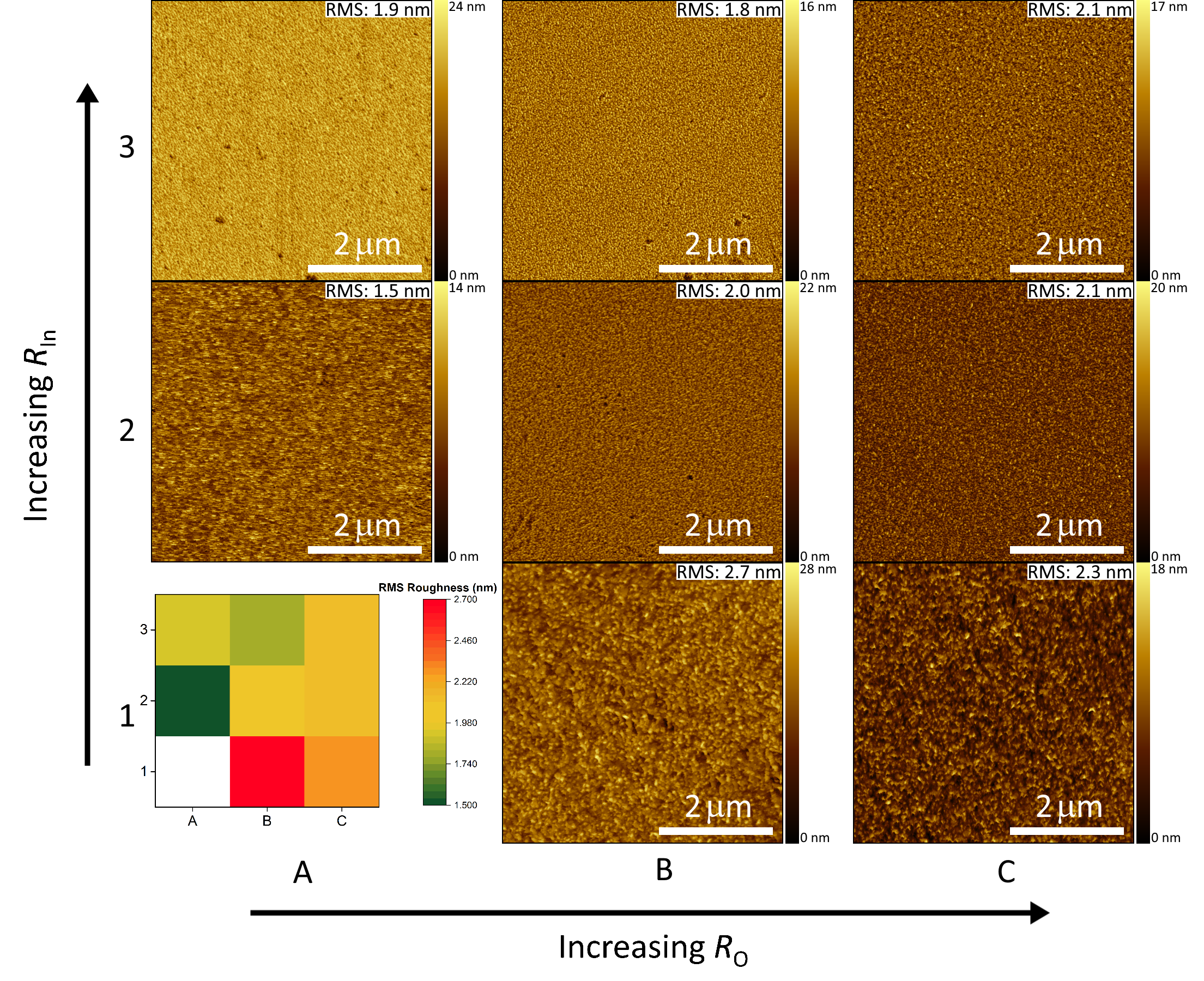}
\captionof{figure}{5 $\upmu$m $\times$ 5 $\upmu$m AFM images of the same samples as in Fig. 2 in the main text. A heat map of RMS roughnesses is included for ease of comparison.}


\renewcommand{\thefigure}{S2}

\includegraphics[width=0.85\textwidth]{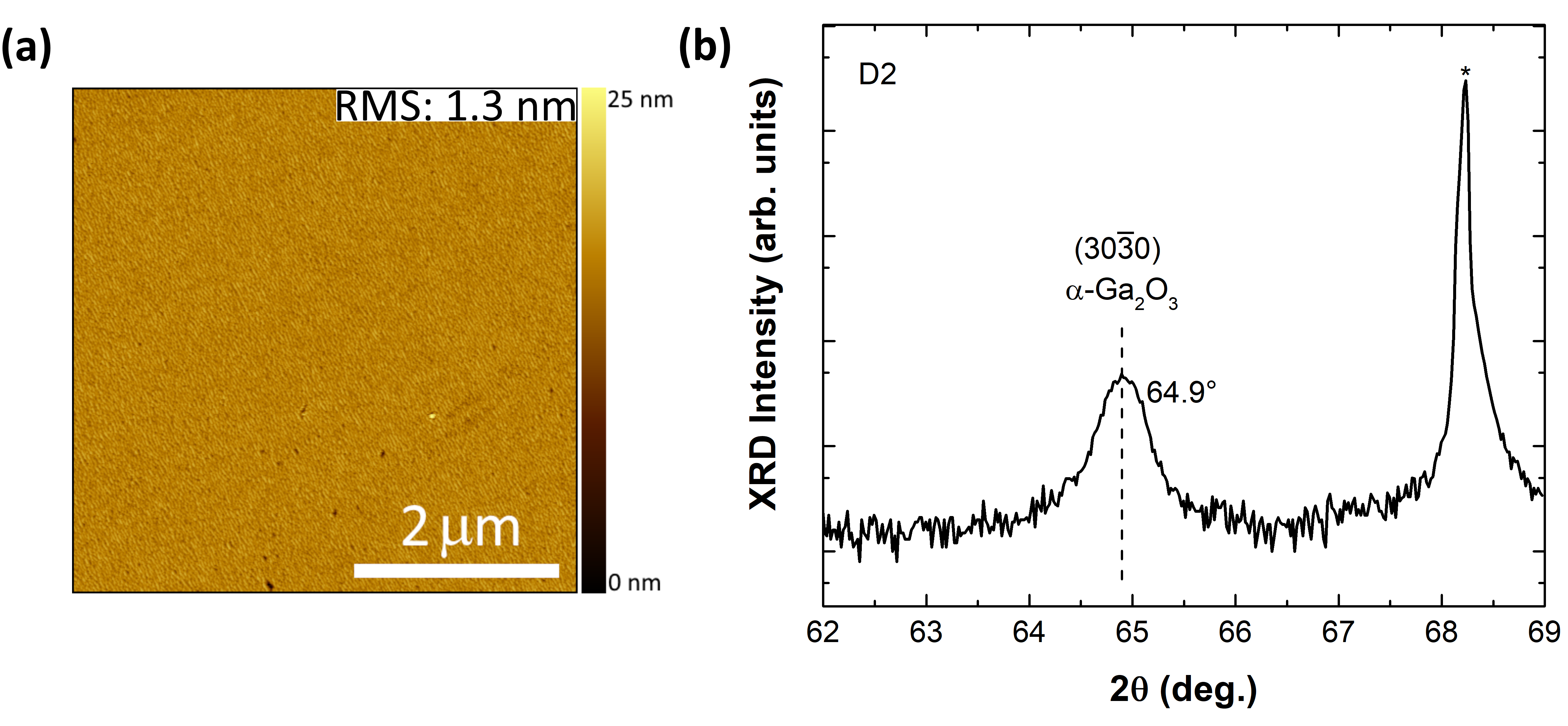}
\captionof{figure}{Sample D2: (a) 5 $\upmu$m $\times$ 5 $\upmu$m AFM image, exhibiting aligned surface faceting, like sample A2. (b) Symmetric HRXRD scan, with the (30$\bar{3}$0) \agox\ reflex positioned at 2$\uptheta$ = 64.9°, i.e. no detectable In concentration by HRXRD. Substrate peak marked by an asterisk.}

\renewcommand{\thefigure}{S3}

\includegraphics[width=0.75\textwidth]{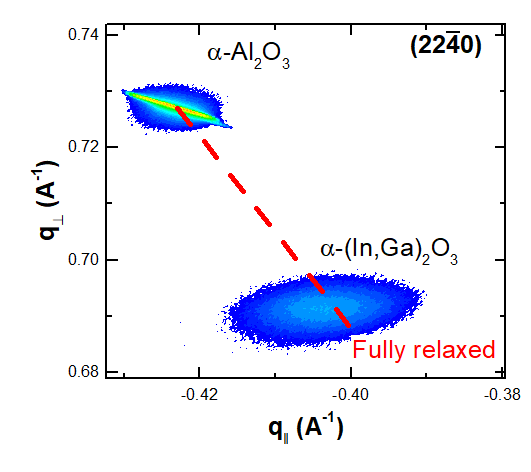}
\captionof{figure}{RSM (q$_{\parallel}$, q$_{\perp}$) around the (22$\bar{4}$0) reflex of sample B2, displaying full relaxation of the \aingox\ film.}

\newpage

\renewcommand{\thefigure}{S4}

\includegraphics[width=0.75\textwidth]{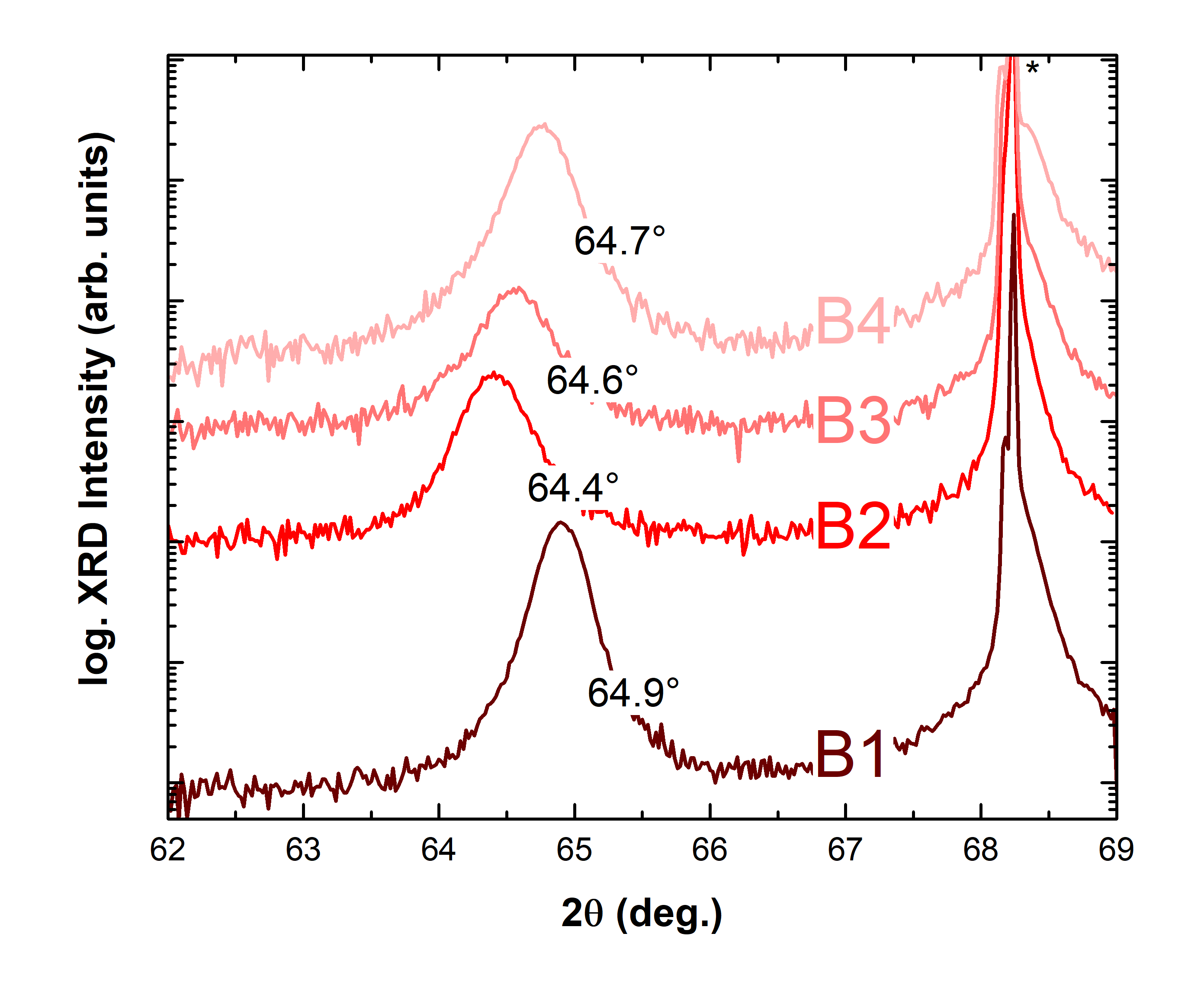}
\captionof{figure}{Symmetric HRXRD \xrdqw~scans of the samples grown at \ogaflux~= 0.80. Substrate peak marked by an asterisk. The (30$\bar{3}$0) \agox~reflex of sample B2 exhibits the greatest shift, and hence the largest In incorporation, $x$ $\approx$ 0.07, of these samples grown at different \ingafluxabb. At higher \ingafluxabb, the (30$\bar{3}$0) \agox~reflex begins returning to the original position of 64.9°, due to a decreasing In concentration. This implies that full In incorporation is reached at $x$ $\approx$ 0.07.}

\renewcommand{\thefigure}{S5}

\includegraphics[width=0.75\textwidth]{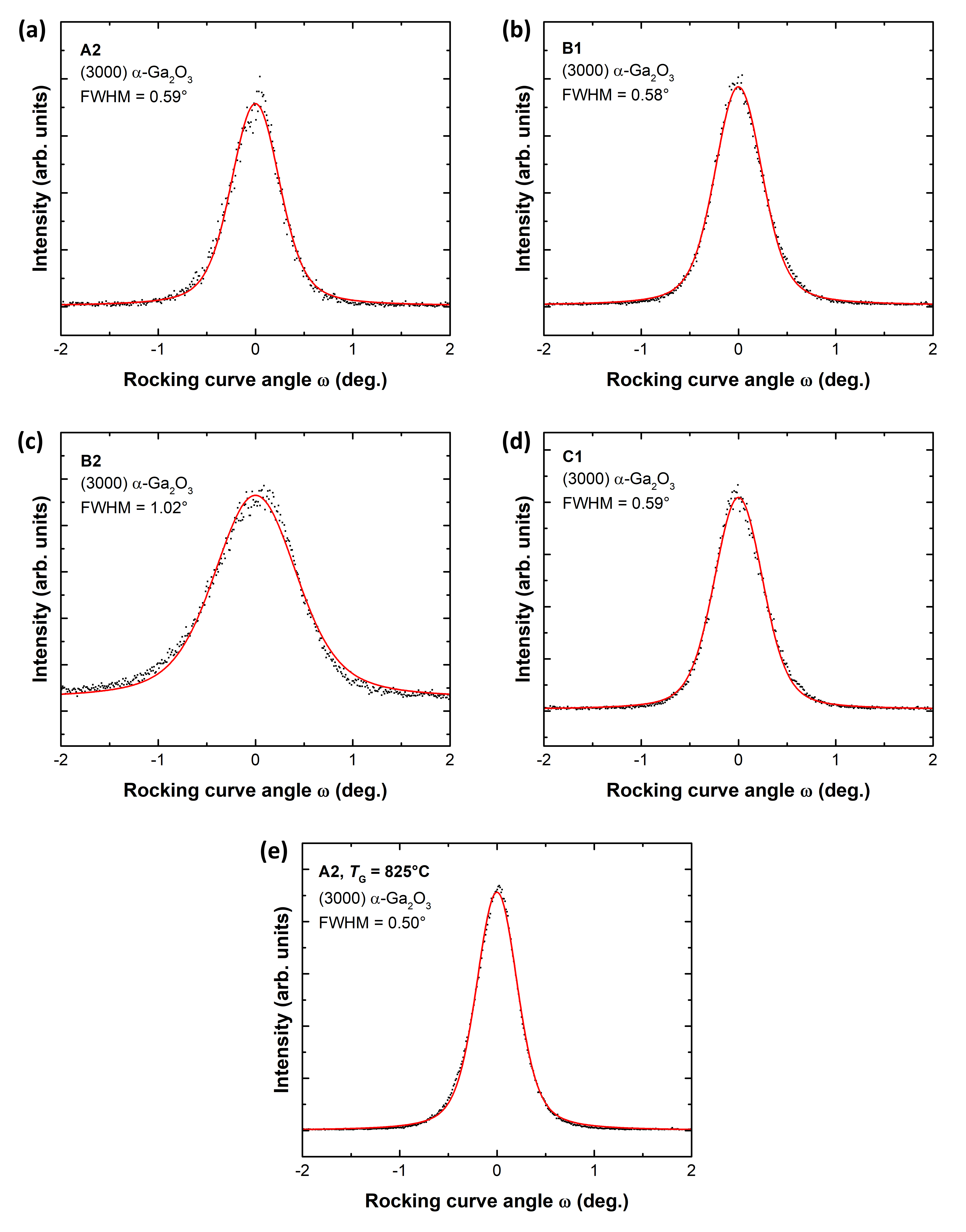}
\captionof{figure}{Rocking curves of samples: (a) A2, (b) B1, (c) B2, (d) C1 and (e) A2 at \tg\ = 825°C. Full width at half maximum (FWHM) values are noted for each sample. The use of In as a precursor during growth has a negligible effect on the FWHM, while In incorporation in the films, such as in sample B2, leads to a broadening of the rocking curve. At higher \tg, the FWHM is reduced considerably.}

\renewcommand{\thefigure}{S6}

\includegraphics[width=0.75\textwidth]{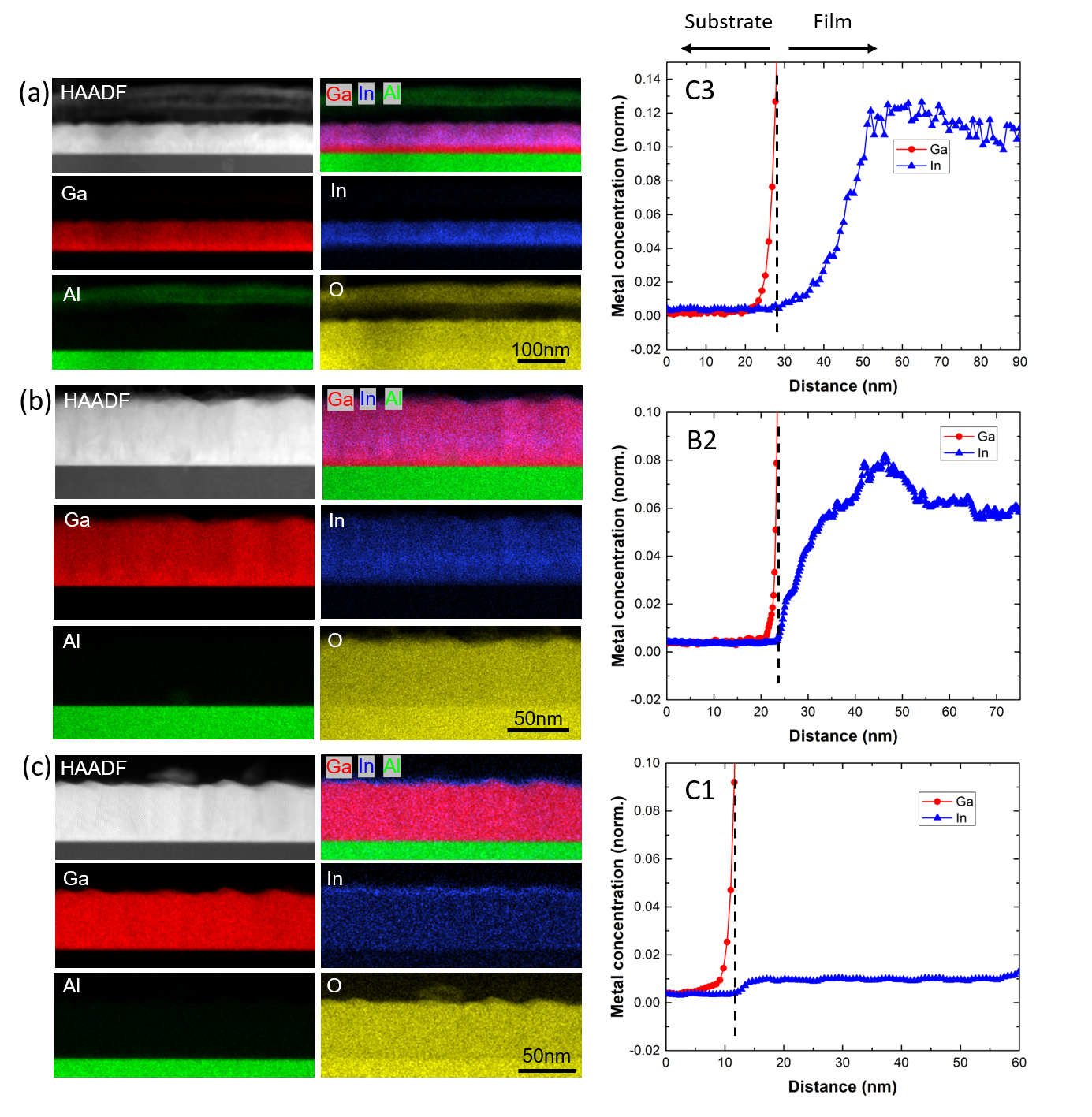}
\captionof{figure}{STEM-EDX composite images and their constituent contributions (Ga = red, In = blue, Al = green, O = yellow), and corresponding metal line scans: (a) C3, (b) B2 and (c) A2. Dashed lines represent the substrate/film interface.}

\newpage

\renewcommand{\thefigure}{S7}

\includegraphics[width=0.75\textwidth]{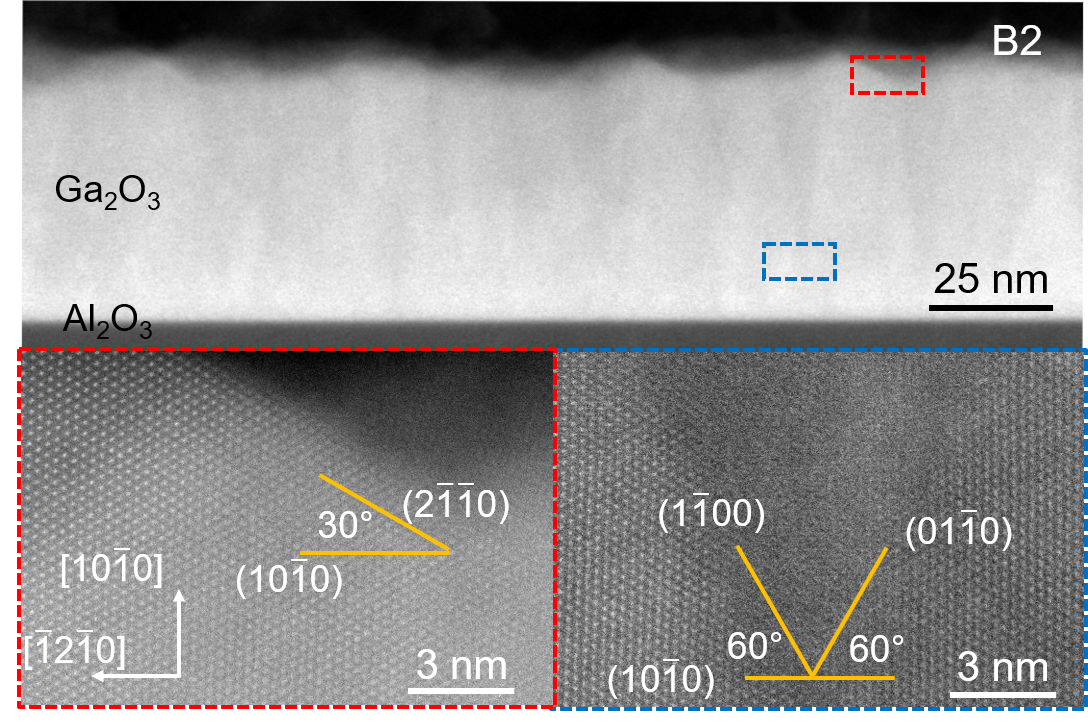}
\captionof{figure}{Sample B2: cross-sectional HAADF overview and magnified images. Two different facets are identified: a-plane (left) with no identifiable growth on top, and m-plane (right) with an unidentified phase growth.}

\renewcommand{\thefigure}{S8}

\includegraphics[width=0.75\textwidth]{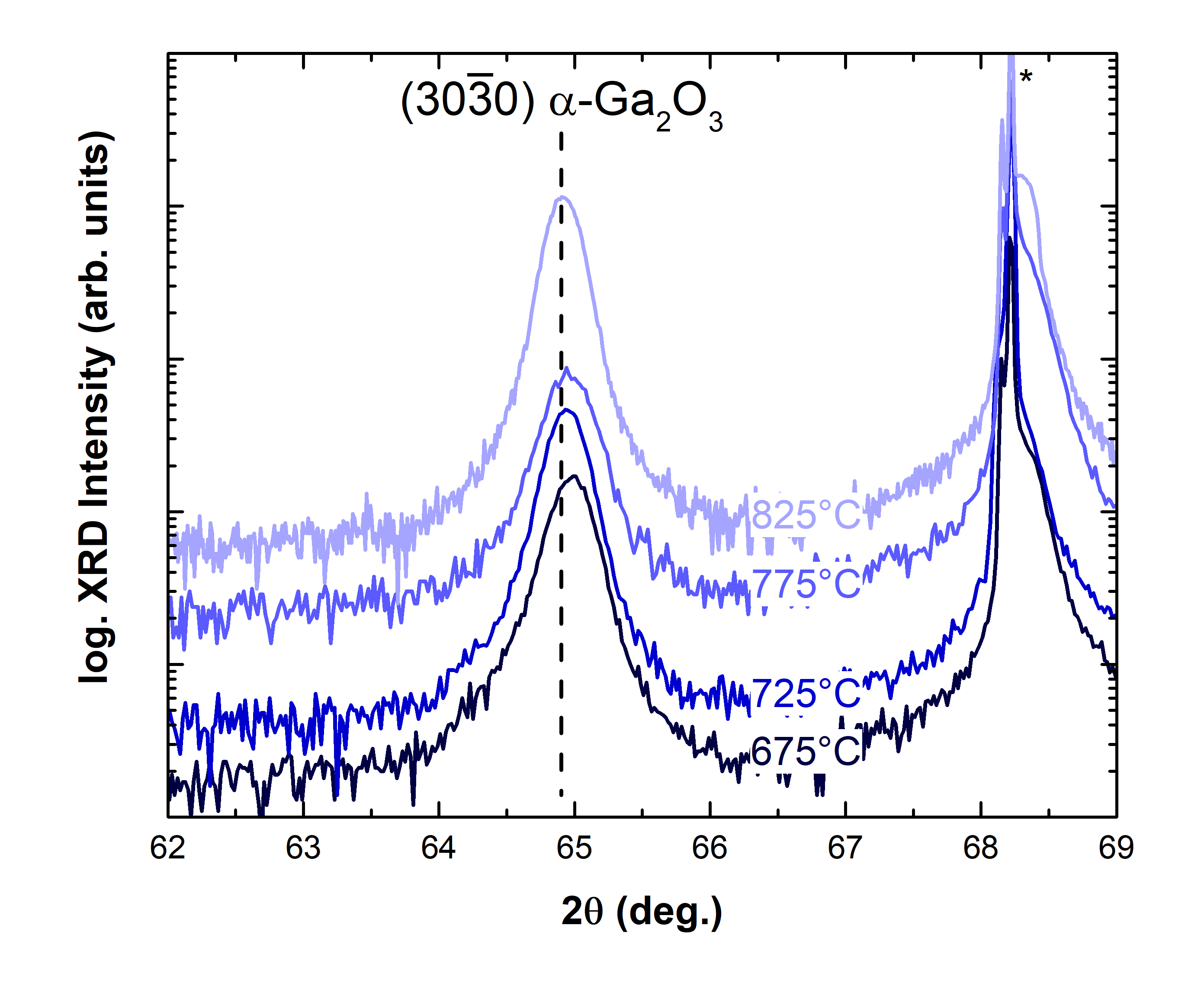}
\captionof{figure}{Symmetric HRXRD \xrdqw~scans of the \tg-dependent layers grown with \ogaflux~and \ingafluxabb~of sample A2. Substrate peak marked by an asterisk. All layers present the expected (30$\bar{3}$0) \agox\ reflex at $\sim$64.9° (dashed line), hence they are \agox, with no detectable In incorporation ($x$ = 0).}

\bibliography{references}